\documentclass[a4paper, twocolumn,superscriptaddress,prd,showpacs,preprintnumbers]{revtex4}
\usepackage{epsf} 
\usepackage{amsmath,color}
\usepackage{amssymb}
\usepackage{epsfig}
\usepackage{latexsym}
\usepackage{amsfonts}
\usepackage{dcolumn}
\usepackage{varioref}
\usepackage{ifthen}
\usepackage{graphicx}
\usepackage{amssymb,amsmath,amsthm}
\begin{document}
\parindent 0mm 
\setlength{\parskip}{\baselineskip} 
\thispagestyle{empty}
\pagenumbering{arabic} 
\setcounter{page}{1}
\mbox{ }
\preprint{UCT-TP-293/13}
\newline
\preprint{MITP/13-003}
\newline
\title{Hadronic Contribution to the muon $g-2$ factor}
\author{S. Bodenstein}
\affiliation{Centre for Theoretical \& Mathematical Physics, University of
Cape Town, Rondebosch 7700, South Africa}
\author{C. A. Dominguez}
\affiliation{Centre for Theoretical \& Mathematical Physics, University of
Cape Town, Rondebosch 7700, South Africa}
\author{K. Schilcher}
\affiliation{Centre for Theoretical \& Mathematical Physics, University of
Cape Town, Rondebosch 7700, South Africa}
\affiliation{Institut f\"{u}r Physik, Johannes Gutenberg-Universit\"{a}t,
Staudingerweg 7, D-55099 Mainz, Germany}
\author{H. Spiesberger}
\affiliation{PRISMA Cluster of Excellence, Institut f\"{u}r Physik,  Johannes Gutenberg-Universit\"{a}t,
Staudingerweg 7, D-55099 Mainz, Germany}

\date{\today}
\begin{abstract}
The lowest order hadronic contribution to the $g-2$ factor of the muon is analyzed in
the framework of the operator product expansion at short distances, and a 
QCD finite energy sum rule designed to quench the role of the $e^+ e^-$ data. This
 procedure reduces the discrepancy between experiment and theory, 
$\Delta a_\mu \equiv a^{EXP}_\mu - a^{SM}_\mu$, from 
$\Delta a_\mu = 28.7 (8.0) \times 10^{-10}$ to $\Delta a_\mu = 19.2 (8.0) \times 10^{-10}$,
i.e. without changing the uncertainty. 

\end{abstract}
%KEYWORDS:  Sum Rules, QCD, quark masses.\\

\maketitle
%\noindent
%

\section{Introduction}
The currently accepted Standard Model (SM) prediction of the muon magnetic anomaly is \cite{davier2011}
%Eq.1
\begin{equation}
a_{\mu}^{\text{SM}}=11\,659\,180.2(4.2)(2.6)(0.2)\times 10^{-10} \;,\label{EQ:SM}
\end{equation}
where the  first error is from the lowest order hadronic contribution, $ a_{\mu}^{\text{HAD,LO}}$,
the second from all other hadronic pieces, and the third is due to all non-hadronic terms. In contrast, the average experimental value is \cite{g2exp1,pdg}
%Eq.2
\begin{equation}
a_{\mu}^{\text{EXP}}=11\,659\,208.9(6.3)\times 10^{-10}\;,\label{EQ:exp}
\end{equation}
leading to a  discrepancy between experiment and theory 
%Eq.3
\begin{equation}
\Delta a_\mu\equiv a_{\mu}^{\text{EXP}}-a_{\mu}^{\text{SM}}=28.7(8.0)\times 10^{-10}\;, \label{EQ:discr}
\end{equation}
which is a  $3.6 \,\sigma$ effect. Another independent analysis of the lowest-order hadronic contribution \cite{hagiwara2011} gives $a_{\mu}^{\text{SM}}=11\,659\,182.8(4.9)\times 10^{-10}$, which corresponds to a $3.3\,\sigma$ discrepancy between experiment and theory.\\

Given that the largest uncertainty in the SM prediction comes from $a_{\mu}^{\text{HAD,LO}}$, and that the  cross-section data from $e^+e^- \rightarrow hadrons$  is used to evaluate this contribution, it is natural to wonder whether the discrepancy $\Delta a_\mu$ could be (partly) due to some issue with the $e^+e^-$ data. The purpose of this paper is to examine the impact on $ a_{\mu}^{\text{HAD,LO}}$ if  the  contribution from the $e^+e^-$ data is quenched by including theoretical input in the framework of the Operator Product Expansion (OPE). This procedure  reduces the discrepancy $\Delta a_\mu$ to  $\Delta a_\mu = 19.2 (8.0) \times 10^{-10}$, i.e. without increasing the error.  In the process of following this procedure it will be shown that there is a clear discrepancy between the OPE based approach and the (unquenched) $e^+e^-$ data. These conclusions depend on two assumptions, viz. (a) the validity of (global) quark-hadron duality, and (b) the dimension $d=2$ term in the OPE is entirely due to quark mass insertions. The validity of these assumptions will be critically discussed. Their impact on the hadronic contribution to the running QED coupling, and hence the SM fitted Higgs boson mass, will be investigated in the Appendix. 
\section{The Method}
\begin{figure}
\centering
\includegraphics[scale=0.65]{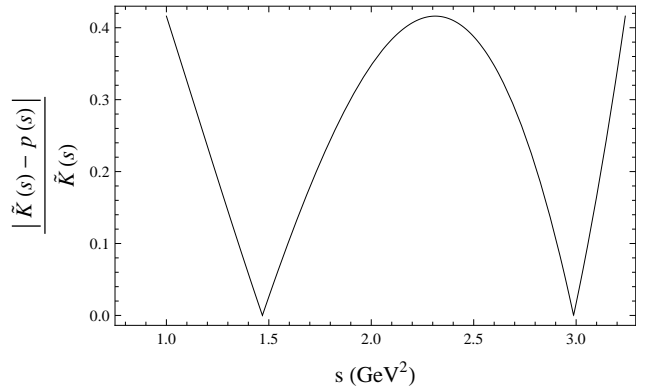}
\caption{{\footnotesize The magnitude of the ratio of integration kernels entering in Eqs. \eqref{EQ:FESR} and \eqref{EQ:dispersion}. The function $p(s)$ is chosen to minimize Eq.\eqref{EQ:MAX}.}} \label{fig:ratio}
\end{figure}

We begin by defining the light-quark electromagnetic current correlator $\Pi(s)$, 
%Eq.4
\begin{eqnarray}
\Pi_{\mu\nu} (q^2) &=& i \int d^4x\,  e^{iqx} \langle 0| T \left(j^{\text{\,EM}}_{\mu}(x)\, j^{\text{\,EM}}_{\nu}(0)\right)|0\rangle \nonumber\\ 
&=& (q_\mu q_\nu - q^2 g_{\mu\nu}) \Pi(q^2)\;, \label{eq:correlator}
\end{eqnarray}
where the electromagnetic current is  $j^{\text{EM}}_\mu(x)=\sum_f Q_f \bar{q}_f(x) \gamma_\mu q_f(x)$, with the sum running over the light quarks $q_f=\{u,d,s\}$, and $Q_f$ are the quark charges. $\Pi(s)$ is analytic in the entire complex squared energy $s$-plane, except for a branch cut along the positive real axis. Unitarity implies the optical theorem $R(s)=12\pi\, \text{Im}\Pi(s)$, where $R(s)$ is the normalized $e^+e^-$ cross-section. The lowest order hadronic contribution to the anomalous magnetic moment of the muon is given by 
%Eq.5
\begin{equation}
a_{\mu}^{\text{HAD},\text{LO}}=\int^{\infty}_{0}\tilde{K}(s)R(s)\,ds \;,\label{EQ:dispersion}
\end{equation}
where $\tilde{K}(s)$ is a well-known weight function. About $92\%$ of the contribution to Eq. \eqref{EQ:dispersion} comes from the low-energy region below $\sqrt{s}=1.8\,\text{GeV}$. Here perturbative QCD (PQCD) is not valid, and cross-section data from $e^+e^-$ experiments is normally used to perform the integration,  followed by PQCD  above $\sqrt{s}=1.8\,\text{GeV}$ \cite{davier2011} and up to the charm-quark threshold. Regarding the heavy-quark contribution, it is even possible to calculate it entirely in  PQCD as shown recently  \cite{bodenstein2012gm2}. Therefore we shall concentrate on the low energy region $\sqrt{s}<1.8\,\text{GeV}$, and discuss how the OPE could be used to quench the contribution of the $e^+e^-$ data. \\
We begin by considering  Cauchy's theorem in the complex $s$-plane
%Eq.6
\begin{equation}
\int^{s_0}_{0}p(s) R(s)\,ds - 6 \pi i \oint_{|s|=s_0}p(s) \Pi(s) \,ds = 0 \;, \label{eq:cauchy}
\end{equation} 
where $p(s)$ is an arbitrary analytic function. Next, we invoke quark-hadron duality  to  replace $\Pi(s)$ by $\Pi_{\text{OPE}}(s)$ in the integral around the circle, where $\Pi_{\text{OPE}}(s)$ is the correlator in the framework of the OPE. Using Eq. \eqref{eq:cauchy}, one can write the low energy piece  of Eq. \eqref{EQ:dispersion}, $\tilde{a}_{\mu}^{\text{HAD},\text{LO}}(s_0)$,  as
%Eq.7
\begin{align}
\tilde{a}_{\mu}^{\text{HAD},\text{LO}}(s_0)=&\int^{s_0}_{0}\bigl[\tilde{K}(s)-p(s)\bigr]R(s)\,ds\nonumber\\
&+6\pi i\oint_{|s|=s_0}p(s)\Pi_{\text{OPE}}(s)\,ds  \;.\label{EQ:FESR}
\end{align}
With the low energy contribution written in this form, one can choose a suitable $p(s)$  to quench the contribution of the  $e^+e^-$ data, for the price of having to perform a contour integral. There are a number of relevant considerations in choosing $p(s)$. First, each power $s^n$ present in  $p(s)$  will involve a dimension $d=2(n-1)$ term in the OPE.  As all terms beyond dimension $d=4$ are currently  poorly known, one should avoid them by considering only kernels of the form $p(s)=a+b s$, thus implying contributions from only dimension $d=2,4$ condensates.  Higher dimensional terms  would only contribute through logarithmic terms at order $\mathcal{O}(\alpha_{s}^{2})$, which are expected to be heavily suppressed \cite{French}. The second consideration is to note that the data in the range $1\,\text{GeV}<\sqrt{s}<1.8\,\text{GeV}$ is particularly badly known. The data in this region have the largest uncertainties, and some of the channels have not even been measured yet (for example the process $e^+e^-\to K\overline{K}\pi$), and their contributions must be estimated using isospin relations. Therefore, we shall  fix the constants $a,b$ in $p(s)$ by demanding that the contribution in the region $1\,\text{GeV}<\sqrt{s}<1.8\,\text{GeV}$ be minimized. In particular,  that the quantity
%Eq.8
\begin{equation}
\text{Max}\left|\frac{\tilde{K}(s)-p(s)}{\tilde{K}(s)}\right|, \ \  (1\,\text{GeV}<\sqrt{s}<1.8\,\text{GeV}) \label{EQ:MAX}
\end{equation}
be minimized. This uniquely fixes $p(s)$ as
%Eq.9
\begin{equation}
p(s)=4.996\times 10^{-9} - 1.432\times 10^{-9} s \;.\label{EQ:fit}
\end{equation}
One can appreciate from Fig. \ref{fig:ratio} that with this choice of $p(s)$  all the data in the interval $\sqrt{s} = (1.0 -1.8) \,\text{GeV}$ is suppressed by a factor of {\bf{at least}} 2.5. It should be stressed that our final results are  highly insensitive to the use of other fitting methods, e.g. least-square fits. As a final comment on the choice of $p(s)$, and in connection with the region $\sqrt{s} < 1.0 \;{\mbox{GeV}}$, the kernel $\tilde{K}(s)$ is approximately proportional to $s^{-2}$, implying that $\tilde{K}(s)$ gets flatter with increasing $s$. The flatter the kernel, the wider the range in $s$ allowing for a linear fit. Quenching data at $s<1\,\text{GeV}^2$  over a sufficiently wide range is not feasible.  The other issue with attempting to suppress the $s<1\,\text{GeV}$ region is that $\tilde{K}(s)-p(s)$ will be larger than $\tilde{K}(s)$ in the poorly known region $\sqrt{s} =1 - 1.8\,\text{GeV}$, thus emphasizing these data. This would defeat the purpose of this method, so that we choose  instead $p(s)$  to satisfy Eq. \eqref{EQ:MAX}. 
%%%%%%%%%%%%%%%%%%%%%%%%%%%%%%%%%%%%%%%%%%%%%%%%%%%%%%%%%%%%%%%%%%%%%%%%%%%%%%%%%%%%%%%%%%%%%%%%%%%%%%%%%%%%%%
\section{OPE} 
In the framework of the OPE the correlator $\Pi(s)$ in Eq. \eqref{eq:correlator} is written as
%Eq.10
\begin{equation}
\Pi(q^2)\,=\, C_0\,\hat{I} \,+\,\sum_{N=0}\;C_{2N+2}(q^2,\mu^2)\;\langle\hat{O}_{2N+2}(\mu^2)   \rangle \;, \label{eq:OPE}
\end{equation}
where $\mu$ is a renormalization scale, and where the Wilson coefficients in this expansion, 
$ C_{2N+2}(q^2,\mu^2)$,  depend on the Lorentz indexes and quantum numbers of $J(x)$ and  of the local gauge invariant operators $\hat{O}_N$ built from the quark and gluon fields. These operators are ordered by increasing dimensionality and the Wilson coefficients, calculable in PQCD, fall off by corresponding powers of $-q^2$. In other words, this OPE achieves a factorization of short distance effects encapsulated in the Wilson coefficients, and long distance dynamics present in the vacuum condensates. The unit operator $\hat{I}$ in Eq. \eqref{eq:OPE} has dimension $d=0$ and $C_0 \hat{I}$ stands for the purely PQCD contribution
%Eq.11
\begin{equation}
\Pi_{\text{PQCD}}(s)=\frac{Q_{T}^{2}}{16\pi^2}\left[\frac{20}{3} -  4 \log\left(-\frac{s}{\mu^2}\right)+\mathcal{O}(\alpha_{s})\right] \;.
\end{equation}
Here $Q_{T}^{2}=\sum_{f=u,d,s} Q_{f}^{2}=2/3$. The perturbative correlator is known up to five-loop order  $\mathcal{O}(\alpha_s^4)$ \cite{chetyrkin1985,gorishnii1991,surguladze1991,chetyrkin1979,dine1979,celmaster1980,baikov2008}. We  use as input the latest PDG world average of the strong coupling \cite{pdg,bethke2012} $\alpha_{s}(M_Z)=0.1184\pm 0.0007$, which we run down to $\mu = M_\tau$ using the Mathematica package \sffamily{Rundec}\normalfont\ \cite{rundec}. Since there are no gauge invariant operators of dimension $d=2$ involving the quark and gluon fields in QCD, it is normally assumed that nonperturbative effects start at dimension $d=4$. This is supported by results from QCD sum rule analyses of $\tau$-lepton decay data \cite{C2}, as well as from lattice QCD (LQCD) \cite{rackow2001} which show no evidence of $d=2$ operators. There is, though, a PQCD term of dimension $d=2$ arising from quark mass corrections, i.e.
%Eq.12 
\begin{equation}
C_{2}\langle\mathcal{O}_{2}\rangle = \sum_{f=u,d,s}  \frac{Q_{f}^{2}}{4\pi^2} \; \frac{\bar{m}_{f}^{2}(\mu)}{s} \,\Bigl[6 + \mathcal{O}(\alpha_{s})\Bigr] \;.\label{EQ:masscorrection}
\end{equation}
Although these mass corrections are negligible for the up- and down-quarks, this is not the case for the strange quark.  We use the three-loop $\mathcal{O}(m_{q}^{2}/s)$ result from \cite{chetyrkin1997}, and the PDG values of the light quark masses \cite{pdg}, i.e.  $\bar{m}_u(2\,\text{GeV})=(2.3\pm 0.7)\,\text{MeV}$, $\bar{m}_d(2\,\text{GeV})=(4.8\pm 0.7)\,\text{MeV}$ and $\bar{m}_s(2\,\text{GeV})=(95\pm 5)\,\text{MeV}$. These values agree within their conservative errors with the most recent determinations from QCD sum rules \cite{mqQCDSR}, as well as from  (LQCD) \cite{FLAG} (for a recent review see \cite{mqREVIEW}).
The dimension $d=4$ term in \eqref{eq:OPE} is given by \cite{chetyrkin1985}
%Eq.13
\begin{align}
&C_{4}\langle\mathcal{O}_{4}\rangle= \frac{1}{s^2} \,\sum_{f=u,d,s} Q_{f}^{2} \Biggl\{\left[\frac{1}{12}-\frac{11}{216}\frac{\alpha_s (\mu)}{\pi}\right]\langle\frac{\alpha_s}{\pi}G^2\rangle \nonumber\\ 
&+\left[2-\frac{2}{3}\frac{\alpha_s(\mu) }{\pi}+\mathcal{O}(\alpha_{s}^{2})\right]\overline{m}_f(\mu) \langle \bar{q_f} q_f\rangle(\mu)\Biggr\}\;.
\end{align}
%Eq.14
%\begin{equation}
%\langle \bar{q} q\rangle = -[267(5)\,\text{MeV}]^3 = - [0.019(4)] {\mbox{GeV}}^3 \;,
%\end{equation}
%at a scale $\mu = 2 \, {\mbox{GeV}}$.
%, and also from LQCD. We will take the recent LQCD value \cite{fukaya2010}
%The strange quark condensate has been determined from QCD sum rules with the result %\cite{sbars}$\langle \bar{s} s\rangle = (0.6 \pm 0.1) \langle \bar{q} q\rangle$, in agreement with %earlier determinations \cite{qbarq2}.\\
The gluon condensate is  known with very large uncertainties, e.g.  LQCD \cite{rackow2005} gives $\langle\frac{\alpha_s}{\pi}G^2\rangle=0.04(1)\,\text{GeV}^4$, while QCD sum rules using the ALEPH $\tau$-decay data yield \cite{C2} $\langle\frac{\alpha_s}{\pi}G^2\rangle=0.01(1)\,\text{GeV}^4$.  There are also determinations based on $e^+e^-$ cross-section data \cite{bodensteinOPE} affected by larger uncertainties.  We shall use here the conservative value 
%Eq.14
\begin{equation}
\langle\frac{\alpha_s}{\pi}G^2\rangle=0.015(15)\,\text{GeV}^4 \;,
\end{equation}
which is consistent with most determinations. The up- (down-) and strange-quark condensates are well known to be determined from the corrections to the Gell Mann-Oakes-Renner relation \cite{qbarq}-\cite{qbarq2}. Using this information it turns out that the uncertainty in the gluon condensate by far dominates over that of the  second term in Eq.(13).
Finally, we also include the QED contribution to the electromagnetic correlator
%Eq.15
\begin{align}
\Pi_{\text{QED}}(s)=\frac{3Q_{T}^{4}}{16\pi^2}\left[\frac{55}{12}-4 \,\zeta_3-\log\left(\frac{-s}{\mu ^2}\right)\right]\frac{\alpha_{\text{EM}}}{\pi} \;,
\end{align} 
where $Q_{T}^{4}=\sum_{f=u,d,s} Q_{f}^{4}=2/9$.
In order to evaluate the integrals involving $R(s)$ in Eq. \eqref{EQ:FESR} or Eq. \eqref{EQ:dispersion}, one needs to collate a complete set of $e^+e^-$ data. We shall use the data set  described in \cite{bodensteinOPE}. The procedure  is to use all available data from \emph{BABAR}, and where these data are not available, use data from the most recent alternative experiment. For channels where there is currently no experimental data, the missing data was estimated using isospin relations. Systematic errors from the same experimental group were conservatively taken to be 100\% correlated. It should also be noted that for $\sqrt{s} > 2\,\text{GeV}$, this data set employs the BES inclusive data \cite{BES2002,BES2009}.

%%%%%%%%%%%%%%%%%%%%%%%%%%%%%%%%%%%%%%%%%%%%%%%%%%%%%%%%%%%%%%%%%%%%%%%%%%%%%%%%%%%%%%%%%%%%%%%%%%%%%%%%%%%%%%%%%%%%%%%%%%%%%%%%%%%%%%%%%%%%%%%%%%%%%%
% Results
%%%%%%%%%%%%%%%%%%%%%%%%%%%%%%%%%%%%%%%%%%%%%%%%%%%%%%%%%%%%%%%%%%%%%%%%%%%%%%%%%%%%%%%%%%%%%%%%%%%%%%%%%%%%%%%%%%%%%%%%%%%%%%%%%%%%%%%%%%%%%%%%%%%%%%
\section{Results} 
\begin{figure}
\centering
\includegraphics[scale=0.65]{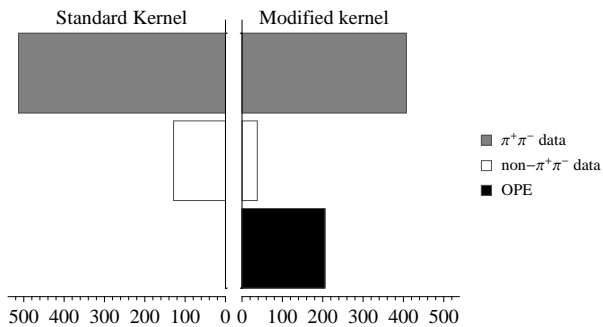}
\caption{{\footnotesize The contribution to $a_{\mu}^{\text{HAD},\text{LO}}$ (in units of $10^{-10}$) from different sources using the standard kernel approach Eq. \eqref{EQ:dispersion} and the modified kernel approach Eq.  \eqref{EQ:FESR}.}}\label{fig:comp}
\end{figure}

Using the standard kernel $\tilde{K}(s)$ in Eq. \eqref{EQ:dispersion}, and our data collection up to $\sqrt{s_0}=1.8\,\text{GeV}$, we find
%Eq.16
\begin{equation}
a_{\mu}^{\text{HAD},\text{LO}}=(640.7\pm 6.5 _{\Delta\text{data}})\times 10^{-10} \;,\label{EQ:result1}
\end{equation}
where the error $\Delta\text{data}$ is the combined statistical and systematic error from the $e^+e^-$ data. 
%This is in agreement within errors with the result from \cite{davier2011} (also at $\sqrt{s_0}=1.8\,\text{GeV}$)
%Eq.17
%\begin{equation}
%a_{\mu}^{\text{HAD},\text{LO}}=(633.9 \pm 4.2)\times 10^{-10} \;.\label{EQ:DAVIER1}
%\end{equation}
Evaluating $a_\mu^{\rm HAD,LO}$ with the quenched kernel, as given in Eq. \eqref{EQ:FESR},
and using Fixed Order Perturbation Theory (FOPT) in the contour integral, with $\sqrt{s_0}=1.8\,\text{GeV}$, and $\mu=M_\tau$, we obtain
%Eq.17
\begin{align}
\tilde{a}_{\mu}^{\text{HAD},\text{LO}}=&(651.0\pm 3.1_{\Delta\text{data}} \pm 1.9_{\Delta\text{conv}}\nonumber\\
&\pm 1.0_{\Delta\alpha_s}\pm 1.3_{\Delta\langle G^2 \rangle})\times
 10^{-10} \nonumber\\
& = (651.0 \pm 4.0) \times 10^{-10}
 \;.\label{EQ:result2}
\end{align}
The uncertainty $\Delta\text{conv}$ is the difference between the $\mathcal{O}(\alpha_{s}^{4})$ and $\mathcal{O}(\alpha_{s}^{3})$ PQCD contributions, $\Delta \alpha_s$ is the error from the strong coupling, and $\Delta \langle G^2 \rangle$  from the uncertainty in the gluon condensate.  The total error above is lower than that from the standard approach in Eq. \eqref{EQ:result1}, primarily due to the different contribution from the data, which amounts to $445.1 \times 10^{-10}$ in Eq. \eqref{EQ:result2}, as opposed to $640.7 \times 10^{-10}$ in Eq. \eqref{EQ:result1}.  Using Contour Improved Perturbation Theory (CIPT) to perform the contour integral in Eq. \eqref{EQ:FESR} we find 
%Eq.18
\begin{align}
\tilde{a}_{\mu}^{\text{HAD},\text{LO}}=&(649.4\pm 3.1_{\Delta\text{data}} \pm 1.4_{\Delta\text{conv}}\nonumber\\
&\pm 0.8_{\Delta\alpha_s} \pm 1.3_{\Delta\langle G^2 \rangle})\times 10^{-10} \nonumber\\
&= (649.4 \pm 3.8) \times 10^{-10}
\;.\label{EQ:result3}
\end{align}
We adopt a conservative approach treating the difference between both methods as another source of uncertainty. Considering the central value to be the average of the FOPT and CIPT results, with an error to be added in quadrature with the other uncertainties, we obtain
%Eq.19
\begin{align}
\tilde{a}_{\mu}^{\text{HAD},\text{LO}}=&(650.2\pm 3.1_{\Delta\text{data}} \pm 1.7_{\Delta\text{conv}} \pm 0.9_{\Delta\alpha_s}\nonumber\\
&\pm 1.3_{\Delta\langle G^2 \rangle} \pm 0.8_{\Delta\text{int}})\times 10^{-10}\nonumber\\
&=(650.2\pm 4.0)\times 10^{-10} \;,\label{EQ:result4}
\end{align}
where $\Delta\text{int}$ is the uncertainty from the integration method (FOPT and CIPT).
This result is more precise than  the standard one, Eq. \eqref{EQ:result1}, due to the suppression of the data contribution achieved by using Eq.\eqref{EQ:FESR}. The relative contributions to $a_\mu^{\rm HAD,LO}$ from different sources is shown in Fig. \eqref{fig:comp}.\\
Regarding the next-to-leading order hadronic contributions to the muon $g-2$, the same $e^+e^-$ data is  used in the standard approach based on Eq. \eqref{EQ:dispersion}.  For instance, using the kernel of \cite{krause} gives $a_{\mu}^{\text{HAD},\text{NLO}}=-10.1(6)\times 10^{-10}$, which has a different sign from the leading-order contribution. Using the same procedure as above, we find the following difference between results from the modified kernel approach, $\tilde{a}_{\mu}^{\text{HAD},\text{NLO}}$, and the standard approach, $a_{\mu}^{\text{HAD},\text{NLO}}$, 
%Eq.20
\begin{equation}
\tilde{a}_{\mu}^{\text{HAD},\text{NLO}}-a_{\mu}^{\text{HAD},\text{NLO}}=-0.2\times 10^{-10}\;, \label{EQ:ho}
\end{equation}
where no error is given, as it is not significant given the uncertainty in Eq. \eqref{EQ:result4}. Adding this next-to-leading order result to Eq. \eqref{EQ:result4} and recalculating the discrepancy, Eq. \eqref{EQ:discr}, gives
%Eq.21
\begin{equation}
\Delta a_\mu\equiv a_{\mu}^{\text{EXP}}-\tilde{a}_{\mu}^{\text{SM}}=19.2(8.0)\times 10^{-10}\;, \label{EQ:discr2}
\end{equation}
which is a lower $2.4 \,\sigma$ effect.

%%%%%%%%%%%%%%%%%%%%%%%%%%%%%%%%%%%%%%%%%%%%%%%%%%%%%%%%%%%%%%%%%%%%%%%%%%%%%%%%%%%%%%%%%%%%%%%%%%%%%%%%%%%%%%
\section{Validity of the method} 
As indicated in the Introduction, the method followed here to determine the leading hadronic contribution to the muon anomaly relies on two basic assumptions, i.e.  (a) (global) quark-hadron duality, and (b) the absence of a dimension $d=2$ term in the OPE, beyond well known PQCD quark mass corrections. We discuss their validity in this section.

The OPE beyond perturbation theory in the deep Euclidean region, involving a parameterization of confinement  in terms of quark and gluon vacuum condensates, is  one of the two pillars of the QCD sum rule method. This allows for analytic determinations of hadronic parameters, as well as of quark masses and the strong coupling \cite{QCDSR_REV}. However, the contour integral in Eq.  \eqref{EQ:FESR} involves an integration of the OPE expression for the correlator near the Minkowski axis, a potential source of difficulties as first pointed out in the pioneering paper \cite{Shankar}. Let us define the difference between the exact electromagnetic correlator and its expression as given by the OPE as 
%Eq.22
\begin{equation}
\Delta_{\text{DV}} = \Pi(s)-\Pi_{\text{OPE}}(s)\;.
\end{equation}
Then the exact result for the contour integral in Eq. \eqref{eq:cauchy} becomes
%Eq.23
\begin{equation}
 \oint_{|s|=s_0}p(s) \Pi(s) ds = \oint_{|s|=s_0}p(s)\left[\Pi_{\text{OPE}}(s)+\Delta_{\text{DV}}\right]\,ds \;. 
\end{equation} 
In order to obtain Eq. \eqref{EQ:FESR} the integral over $\Delta_{\text{DV}}$ was assumed to vanish, in other words no violation of (global) quark-hadron duality. This is supported by a model independent  analysis \cite{dominguez2009} based on the vector and axial-vector spectral functions from hadronic $\tau$-decays as measured by the ALEPH collaboration \cite{ALEPH}. However, this is currently a contentious issue, as there are model dependent claims to the contrary. In order to determine the impact of potential duality violations on our results we consider first a recent quantitative model for $\text{Im}\,\Delta_{\text{DV}}(s)$ \cite{cata2009}, which gives
%Eq.24
\begin{align}
\frac{1}{\pi}\text{Im}\,\Delta_{\text{DV}}(s)&=\sum_{i=u,d,s}Q_{i}^{2}\Bigl[\frac{5}{6}\kappa_V e^{-\gamma_V s}\sin(\alpha_V+\beta_V s)\nonumber\\
&+\frac{1}{6}\kappa_V e^{-\gamma_V s}\sin(\alpha'_V+\beta_V s)\Bigl] \;\;,
\end{align}
where the values of the various constants may be found in \cite{cata2009}. Using this model we obtain
%Eq.25
\begin{align}
6\pi i \oint_{|s|=s_0}p(s)\,\Delta_{\text{DV}}\,ds&=-12\pi\int^{\infty}_{s_0} p(s) \text{Im}\,\Delta_{\text{DV}}(s)\,ds\nonumber\\ 
&= - 0.59(59)\times 10^{-10} \;.
\end{align}
The error above stems from the uncertainties in the constants $\kappa_V,\alpha'_V,\alpha_V,\gamma_V,\beta_V$ \cite{cata2009}. Since no error correlations were given in \cite{cata2009} we added them in quadrature. This result is obviously consistent with no duality violation. In any case, it provides evidence that a potential duality violation is not large enough to make Eq. \eqref{EQ:result1} consistent with Eq. \eqref{EQ:result2}. \\ 
We consider next a test involving large variations in the value of $s_0$, for a given integration kernel $p(s)$. This kernel should remain fixed, as otherwise there would be changes to the data subsets being quenched, thus preventing any conclusion. In general, one expects the size of any potential duality violation to decrease with increasing $s_0$. In fact, recent high-precision BES inclusive data \cite{BES2009} in the range  $\sqrt{s}= (2.60\, - 3.65)\; \text{GeV}$  are already consistent with PQCD.
%FIG.3
\begin{figure}
\centering
\includegraphics[scale=0.65]{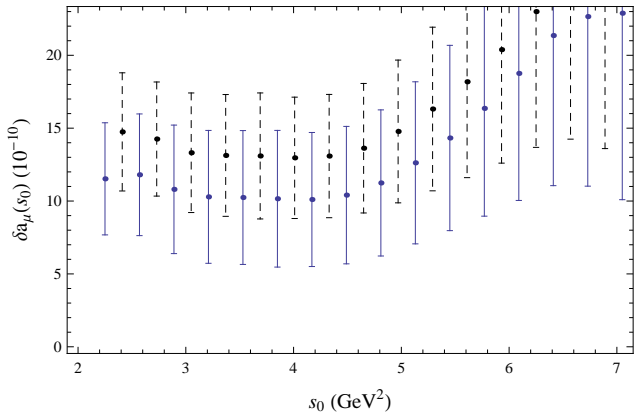}
\caption{{\footnotesize The difference  $\delta a_\mu\equiv\tilde{a}_{\mu}^{\text{HAD},\text{LO}}-a_{\mu}^{\text{HAD},\text{LO}}$ as a function of $s_0$, where $\tilde{a}_{\mu}^{\text{HAD},\text{LO}}$ is given in Eq. \eqref{EQ:FESR}, and $a_{\mu}^{\text{HAD},\text{LO}}$ in Eq. \eqref{EQ:dispersion}.
$\Delta a_\mu$ should vanish according to Cauchy's theorem. The contour integral in Eq. \eqref{EQ:FESR} was evaluated using FOPT. The dashed (solid) line is the result from using the 2010 KLOE \cite{kloe2010} (2009 BABAR \cite{babar2012}) data  in the $e^+e^-\to \pi^+\pi^-$ channel.}}\label{fig:diff}
\end{figure}
In Fig. \ref{fig:diff}
we show the difference $\delta a_\mu\equiv\tilde{a}_{\mu}^{\text{HAD},\text{LO}}-a_{\mu}^{\text{HAD},\text{LO}}$ as a function of $s_0$, where $\tilde{a}_{\mu}^{\text{HAD},\text{LO}}$ is given in Eq. \eqref{EQ:FESR}, and $a_{\mu}^{\text{HAD},\text{LO}}$ in Eq. \eqref{EQ:dispersion}. Clearly,  this difference exhibits a minimum at $\sqrt{s_0}=1.8\,\text{GeV}$, and changes around this value would only increase $\delta a_\mu$, thus reducing the current discrepancy, Eq. \eqref{EQ:discr}. It should be emphasized that we have used the available data on $R(s)$ from BES to obtain the results shown in Fig.3. These data lie slightly above the PQCD prediction for $\sqrt{s} = 2.2 -2.6 \;{\mbox{GeV}}$ (see Fig.8 in \cite{davier2011}). Since our modified integration kernel emphasizes, instead of quenching the impact of data in this energy range, the result for $\delta a_\mu$ increases. This means that the discrepancy between data and theory for $g-2$ is further reduced. We have checked that by using PQCD in a range above a given value of $s_0$, the difference $\delta a_\mu(s)$ becomes frozen for $s > s_0$. 
These results also highlight the current tension  between the OPE and the $e^+e^-$ data, and they imply a problem  with the OPE and/or with the $e^+e^-$ data. This problem can hardly be blamed on duality violations, as $\delta a_\mu$ grows with increasing $s_0$.\\
As it can be appreciated from Fig. \ref{fig:diff}, the kernel Eq. \eqref{EQ:fit}
causes $\delta a_\mu$ to rapidly increase  above $s_0 \simeq 5\; {\mbox{GeV}}^2$, although results are still consistent within the large, and also increasing errors. This is due to the specific functional form  of $p(s)$, Eq.\eqref{EQ:fit}, which dominates over the kernel $\tilde{K}(s)$ at large $s$. Our
 result for $\delta a_\mu$ could superficially appear in contradiction with  those of \cite{hagiwara2011} where, within large errors, no discrepancy was found using kernels of the form
%Eq.26
\begin{equation}
p_{m,n}(s,s_0) = \left(1 - \frac{s}{s_0}\right)^m \, \left(\frac{s}{s_0}\right)^n \;, \label{EQ:H}
\end{equation}
where the integers $m,n$ take a variety of values in the range $m,n = 0-1$. Notice that these kernels are of order ${\cal{O}}(1)$ compared with our kernel, Eq.(9), which is of order ${\cal{O}}(10^{-9})$. This overall factor, though, is irrelevant for the uncovering of the discrepancy.
In Table 1 we show typical results for $\delta a_\mu$ using our method, but replacing Eq. \eqref{EQ:fit} with the above kernel. For certain combinations of the powers $m,n$ the kernel Eq. \eqref{EQ:H} is unable to unveil the underlying discrepancy. However, for the combination $m=1, n=0$, Eq. \eqref{EQ:H} becomes similar to our kernel, Eq. \eqref{EQ:fit} (up to an overall factor), thus partially uncovering the tension. Since different kernels emphasize different regions of the data, some kernels are better suited than others to uncover this discrepancy.\\
%Table1
\begin{table}
\begin{ruledtabular}
\begin{tabular}{cccc}
\cline{1-4}
\noalign{\smallskip}
 $m$  & $n$ & $\sqrt{s_0}$  & $\delta a_\mu (s_0)$ \\
 &&& ($10^{-2}$)\\
\hline
\noalign{\smallskip}

$0$ & $0$ & $2.6\, {\mbox{GeV}}$ & $- 17 (52)$ \\
$1$ & $0$ & $2.6\, {\mbox{GeV}}$ & $ 12 (21)$ \\
$0$ & $0$ & $2.0\, {\mbox{GeV}}$ & $ 26 (32)$ \\
$1$ & $0$ & $2.0\, {\mbox{GeV}}$ & $ 18 (11)$ \\

\end{tabular}
\caption{\footnotesize{The difference $\delta a_\mu\equiv\tilde{a}_{\mu}^{\text{HAD},\text{LO}}-a_{\mu}^{\text{HAD},\text{LO}}$ using
 the integration kernel \cite{hagiwara2011} Eq. \eqref{EQ:H}, instead of our kernel, Eq. \eqref{EQ:fit}. Except possibly for the case $m=1$, $n=0$, 
this kernel is unable to uncover the discrepancy shown in Fig. 3.}}
\end{ruledtabular}
\end{table}

We now examine the second assumption, i.e. the  absence of a dimension $d=2$ gauge non-invariant term in the OPE. There is evidence in support of this assumption from  LQCD  \cite{rackow2001}, as well as from QCD sum rule analyses \cite{C2,dominguez2009} of the $\tau$-decay data from the ALEPH collaboration \cite{ALEPH}. However, this assumption could be somewhat relaxed, as we only require the contribution of a  $d=2$ term to be positive for our conclusions to continue to hold (a positive contribution to Eq. \eqref{EQ:result4} implies a negative value of the $d=2$ term in the OPE). There is a model for an effective $d=2$ term proposed long ago \cite{narison1999} in connection with a so-called phenomenological gluon mass given by
%Eq.27 
\begin{equation}
C_2\langle \mathcal{O}_2\rangle =\sum_{f=u,d,s} Q_{f}^{2}\frac{1}{16\pi^2}\frac{\alpha_s}{\pi}\lambda^2\left(\frac{128}{3}-32\zeta(3)\right) \;,\label{EQ:tachyonic}
\end{equation}
where $\lambda^2$ is a tachyonic gluon mass ($\lambda^2<0$), and $\zeta(n)$ is the Riemann zeta function. The gluon mass was estimated in the range \cite{narison1999} $-0.085\,\text{GeV}^2<\frac{\alpha_s}{\pi}\lambda^2<-0.034\,\text{GeV}^2$. If this term is included in the OPE Eq. \eqref{eq:OPE}, our results Eqs. \eqref{EQ:result2}-\eqref{EQ:result4} would {\bf{increase}} by an amount $(3.6 - 8.9)\times 10^{-10}$, significantly decreasing the discrepancy $\Delta a_\mu$, Eq. \eqref{EQ:discr}.   

%%%%%%%%%%%%%%%%%%%%%%%%%%%%%%%%%%%%%%%%%%%%%%%%%%%%%%%%%%%%%%%%%%%%%%%%%%%%%%%%%%%%%%%%%%%%%%%%%%%%%%%%%%%%%%%%%%%%%%%%%%%%%%%%%%%%%%%%%%%%%%%%%%%%%%
% Conclusion
%%%%%%%%%%%%%%%%%%%%%%%%%%%%%%%%%%%%%%%%%%%%%%%%%%%%%%%%%%%%%%%%%%%%%%%%%%%%%%%%%%%%%%%%%%%%%%%%%%%%%%%%%%%%%%%%%%%%%%%%%%%%%%%%%%%%%%%%%%%%%%%%%%%%%%
\section{Conclusion} 
In this paper we examined within the SM the possibility of reducing the discrepancy between the experimental and the theoretical value of the muon anomaly, Eq. \eqref{EQ:discr}, which stands as a $3.6\; \sigma$ effect. We made two basic assumptions, (a) the validity of (global) quark-hadron duality, and (b) the absence of a gauge non-invariant dimension $d=2$ term in the OPE, thus achieving a quenching of the contribution of the $e^+e^-$ data in the region $1\,\text{GeV}<\sqrt{s}<1.8\,\text{GeV}$. Using our $e^+ e^-$ data compilation, and Eq. \eqref{EQ:FESR}, we achieved a reduction of the current discrepancy, Eq. \eqref{EQ:discr}, by $9.5\times 10^{-10}$, leading to Eq. \eqref{EQ:discr2}, which is now a lower $2.4\, \sigma$ effect. This reduction would increase further by increasing $s_0$, or by a different choice of data in the two-pion channel, e.g. the 2010 KLOE data \cite{kloe2010}. In this respect, the most recent KLOE analysis \cite{kloe2012} is fully consistent with \cite{kloe2010}, and has similar uncertainties.\\
In addition, we found a clear tension between our $e^+e^-$ data compilation and the OPE, as shown in Fig. \ref{fig:diff}. 
%Quantitatively, taking the difference between Eq. \eqref{EQ:result4} and Eq. \eqref{EQ:result1}, we find
%Eq.28
%\begin{equation}
%\tilde{a}_{\mu}^{\text{HAD,LO}}-a_{\mu}^{\text{HAD,LO}}=8.3(3.5)\times 10^{-10} \;,
%\end{equation}
%which deviates from 0 by $2.4\sigma$.
This suggests a cross-section deficit in the $e^+ e^-$ data. We  conclude that neither potential duality violations nor a hypothetical gauge non-invariant dimension $d=2$ term in the OPE are likely to account for this current discrepancy between the OPE and the $e^+e^-$ data. 
\section {Appendix}
%\section{The hadronic contribution to the running QED coupling}
The $e^+e^-$ cross-section data used for the determination of the hadronic contribution to the magnetic moment of the muon is also used for the calculation of the hadronic contribution to the running QED coupling $\Delta \alpha_{\text{HAD}}(M_Z)$, which is given by
%Eq.28
\begin{equation}\label{EQ:alpha}
\Delta \alpha_{\text{HAD}}(M_Z)=\frac{\alpha_{\text{em}} M_{Z}^{2}}{3\pi}P\int^{\infty}_{0}\frac{R(s)}{s(M_{Z}^{2}-s)}ds \;.
\end{equation}
$\Delta \alpha_{\text{HAD}}(M_Z)$ is an important input to Standard Model fits to the Higgs boson mass. Currently, a value of $M_H=91^{+30}_{-23}\,\text{GeV}$ \cite{davier2011} is obtained from such fits, compared to the mass of a potential Higgs boson detected by the ATLAS Collaboration with $M_H=125.3(0.6)\,\text{GeV}$ \cite{atlas} or by the CMS Collaboration with $M_H=126.0(0.6)\,\text{GeV}$ \cite{cms}. It has been noted \cite{marciano2008} that modifications to the $e^+e^-$ cross-section data that reduce the discrepancy in the muon $g-2$, i.e. $\Delta a_\mu$, also increases $\Delta \alpha_{\text{HAD}}(M_Z)$, and hence decreases the fitted Higgs boson mass. This increases the tension between the measured and fitted Higgs boson masses. It is interesting in this context to examine the effect of reducing the contribution of the $e^+e^-$ data to $\Delta \alpha_{\text{HAD}}(M_Z)$ in the $1\,\text{GeV}<\sqrt{s}<1.8\,\text{GeV}$ region  using an analogue of Eq. \eqref{EQ:FESR}. All that needs to be changed is the integration kernel $\tilde{K}(s)\to \frac{\alpha M_{Z}^{2}}{3\pi}\frac{1}{s(M_{Z}^{2}-s)}$ in Eq. \eqref{EQ:dispersion}. We find that the linear integration kernel $p_\alpha(s)=0.0008490 - 0.0002002 s$  quenches all the data contribution in the interval $1\,\text{GeV}<\sqrt{s}<1.8\,\text{GeV}$ by {\bf{at least}} a factor of 6.25. The greater suppression in this case is due to the kernel in Eq. \eqref{EQ:alpha} behaving as $\sim s^{-1}$ in this region, rather than as $\sim s^{-2}$ for the $g-2$ kernel. It is thus less sloped, and hence it is better fitted by the  linear function $p_\alpha(s)$. Denoting by $\delta \alpha_{\text{HAD}}$  the difference between the modified kernel result and the standard result we find, using $\sqrt{s_0}=1.8\,\text{GeV}$,
%Eq.29
\begin{equation}
\delta \alpha_{\text{HAD}}=\left\{\begin{matrix} 
+1.9\times 10^{-4} \ \ \ \ (\text{FOPT}) \\
+1.5\times 10^{-4} \ \ \ \ \ (\text{CIPT}) 
\end{matrix}\right.
\end{equation}
This increase in $\Delta \alpha_{\text{HAD}}(M_Z)$ decreases  the Higgs boson mass from $M_H=91^{+30}_{-23}\,\text{GeV}$ \cite{davier2011} to approximately $M_H\sim(83-85)\,\text{GeV}$. 
\section{Acknowledgments} 
This work has been supported in part by the Alexander von Humboldt Foundation.
%%%%%%%%%%%%%%%%%%%%%%%%%%%%%%%%%%%%%%%%%%%%%%%%%%%%%%%%%%%%%%%%%%%%%%%%%%%%%%%%%%

\end{document}